\documentclass[pra,aps,showpacs,twocolumn]{revtex4}
\usepackage{graphicx}
\usepackage{latexsym}
\usepackage{amsmath}
\usepackage[dvipdfm]{hyperref}

\begin{document}

\title{Global Quantum discord of multi-qubit states}

\author{Jianwei Xu}
\email{xxujianwei@yahoo.cn}
\affiliation{Key Laboratory for Radiation Physics and Technology, Institute of Nuclear Science and Technology, Sichuan University,
Chengdu 610065, China}

\date{\today}

\begin{abstract}
Global quantum discord (GQD), proposed by Rulli and Sarandy
[Phys. Rev. A \textbf{84}, 042109 (2011)], is a generalization of quantum discord to
multipartite states. In this paper, we provide an equivalent expression for
GQD, and obtain the analytical expressions of GQD for two classes of
multi-qubit states. The phenomena of sudden transition and freeze of GQD are also discussed.
\end{abstract}


\maketitle

\section{Introduction}
Quantifying the multipartite quantum correlations is a very challenging and
still largely open question \cite{Coffman2000,Osborne2006,Zhou2006,Kaszlikowski2008,Rulli2011}. For bipartite case, entanglement and quantum
discord have been widely accepted as two fundamental tools to quantify
quantum correlations \cite{Horodecki2009,Modi2011}, and quantum discord captures more quantum
correlations than entanglement in the sense that a separable state may have
nonzero quantum discord. Generalizations of bipartite quantum discord to
multipartite states have been considered in different ways \cite{Modi2011-2}. In \cite{Rulli2011}, Rulli
and Sarandy proposed a measure for multipartite quantum correlations, called
global quantum discord (GQD), which can be seen as a generalization of
bipartite quantum discord \cite{Zurek2001,Vedral2001} to multipartite states. GQD is always nonnegative and its
use is illustrated by the Werner-GHZ state and the Ashkin-Teller model \cite{Rulli2011}.

In this paper, we provide an equivalent expression for GQD, and give an
interpretation of GQD (Sec.III). We derive the analytical expressions of GQD
for two classes of multi-qubit states (Sec.IV), these results generalize the
earlier results \cite{Zurek2001,Luo2009}. The phenomena of sudden transition and freeze of GQD
are also discussed (Sec.V). For clarity of reading, we first recall the
definition of GQD proposed in \cite{Rulli2011} (Sec.II).

\section{global quantum discord (GQD)}

We briefly review the definition of GQD proposed in \cite{Rulli2011}.

Consider two systems $A_{1}$ and $A_{2}$ (each of
them is of finite dimension), the symmetric quantum discord of a state $\rho
_{A_{1}A_{2}}$ of the composite systems $A_{1}A_{2}$ is
\begin{eqnarray}
D(\rho _{A_{1}A_{2}})=\min_{\Phi }[I(\rho _{A_{1}A_{2}})-I(\Phi
_{A_{1}A_{2}}(\rho _{A_{1}A_{2}}))].
\end{eqnarray}
In Eq.(1),
\begin{eqnarray}
I(\rho _{A_{1}A_{2}})=S(\rho _{A_{1}})+S(\rho _{A_{2}})-S(\rho
_{A_{1}A_{2}}),
\end{eqnarray}
is the mutual information of $\rho _{A_{1}A_{2}}$, $min$ is taken over all
locally projective measurements performing on AB, $\Phi _{(\cdot )}$ denotes a locally
projective measurement performing on the system $(\cdot )$, $S(\cdot )$ is
the Von Neumann entropy, and $\rho _{A_{1}}$, $\rho _{A_{2}}$ are reduced
states of $\rho _{A_{1}A_{2}}$.

$D(\rho _{A_{1}A_{2}})$ is a natural extension of the original definition of
quantum discord which defined over all projective measurements performing
only on $A_{1}$ or $A_{2}$ \cite{Zurek2001,Vedral2001}.

Since the mutual information $I(\rho _{A_{1}A_{2}})$ can be expressed by the
relative entropy
\begin{eqnarray}
I(\rho _{A_{1}A_{2}})=S(\rho _{A_{1}A_{2}}||\rho _{A_{1}}\otimes \rho
_{A_{2}}),
\end{eqnarray}
hence, Eq.(1) can also be recasted as
\begin{eqnarray}
D(\rho _{A_{1}A_{2}})=\min_{\Phi }[S(\rho _{A_{1}A_{2}}||\rho
_{A_{1}}\otimes \rho _{A_{2}})   \nonumber \\
-S(\Phi _{A_{1}A_{2}}(\rho _{A_{1}A_{2}})||\Phi _{A_{1}}(\rho
_{A_{1}})\otimes \Phi _{A_{2}}(\rho _{A_{2}}))].
\end{eqnarray}
Note that the relative entropy of state $\rho $ with respect to state $%
\sigma $ ($\rho $ and $\sigma $ lie on the same Hilbert space) is defined as
\begin{eqnarray}
S(\rho ||\sigma )=tr(\rho \log _{2}\rho )-tr(\rho \log _{2}\sigma ).
\end{eqnarray}
Further, Eq.(1) can also be rewritten as
\begin{eqnarray}
D(\rho _{A_{1}A_{2}})=\min_{\Phi }[S(\rho _{A_{1}A_{2}}||\Phi
_{A_{1}A_{2}}(\rho _{A_{1}A_{2}}))  \nonumber \\
-S(\rho _{A_{1}}||\Phi _{A_{1}}(\rho _{A_{1}}))-S(\rho _{A_{2}}||\Phi
_{A_{2}}(\rho _{A_{1}}))].
\end{eqnarray}
The definition of GQD is a generalization of bipartite symmetric quantum
discord. Consider $N$ $(2\leq N<\infty )$ systems $A_{1}$, $A_{2}$ , ... , $%
A_{N}$ (each of
them is of finite dimension),  the GQD of state $\rho
_{A_{1}A_{2}...A_{N}}$ on the composite system $A_{1}A_{2}...A_{N}$ is defined as \cite{Rulli2011}
\begin{eqnarray}
&& \ \ D(\rho _{A_{1}A_{2}...A_{N}})   \nonumber  \\
&&=\min_{\Phi }[S(\rho
_{A_{1}A_{2}...A_{N}}||\Phi _{A_{1}A_{2}...A_{N}}(\rho
_{A_{1}A_{2}...A_{N}}))  \nonumber  \\
&& \ \ \ \ \ \ \ \ \ \   -\sum_{j=1}^{N}S(\rho _{A_{j}}||\Phi _{A_{j}}(\rho _{A_{j}}))].
\end{eqnarray}
It has been proved that $D(\rho _{A_{1}A_{2}...A_{N}})\geq 0$ for any state $%
\rho _{A_{1}A_{2}...A_{N}}$ \cite{Rulli2011}. Also, it is easy to see that $D(\rho
_{A_{1}A_{2}...A_{N}})$ keeps invariant under any locally unitary
transformation.

\section{An equivalent expression for global quantum discord}

In this section, we provide an equivalent expression for GQD. We first state two mathematical facts as the lemmas below.

\emph{Lemma 1}. For any square matrix (with finite dimension) $A$,  let $%
\overline{A}$ be the matrix whose diagonal elements are the same with $A$,
and other elements are zero. $B$ and $\overline{B}$ are defined similarly. Then
\begin{eqnarray}
&& \ \  tr(A\overline{B})=tr(\overline{A} \ \overline{B}),    \\
&&  tr(Af(\overline{B}))=tr(\overline{A}f(\overline{B})),
\end{eqnarray}
where $f()$ is any function.

\emph{Lemma 2}. Let $\rho _{A_{1}A_{2}...A_{N}}$ be a state on Hilbert space $%
H_{12...N}$, $\rho _{A_{1}}$, $\rho _{A_{2}}$, ... , $\rho _{A_{N}}$ be the
reduced states of $\rho _{A_{1}A_{2}...A_{N}}$ on Hilbert spaces $H_{1}$, $%
H_{2}$, ... , $H_{N}$, respectively. Suppose $\sigma _{A_{1}}$, $\sigma
_{A_{2}}$, ..., $\sigma _{A_{N}}$ are states on $H_{1}$, $H_{2}$, ... , $%
H_{N}$, respectively. Then it holds that
\begin{eqnarray}
&& \ \ \ tr[\rho _{A_{1}A_{2}...A_{N}}\log _{2}(\sigma _{A_{1}}\otimes \sigma
_{A_{2}}\otimes ...\otimes \sigma _{A_{N}})]     \nonumber \\
&&  =\sum_{i=1}^{N}tr_{i}[\rho _{A_{i}}\log _{2}\sigma _{A_{i}}].
\end{eqnarray}
\emph{Proof}. We only prove the case of $N=2$, the proof of $N>2$ is similar. When $%
N=2$, we need to prove
\begin{eqnarray}
&& \ \ \   tr[\rho _{A_{1}A_{2}}\log _{2}(\sigma _{A_{1}}\otimes \sigma _{A_{2}})]    \nonumber \\
&&         =tr_{1}[\rho _{A_{1}}\log _{2}\sigma _{A_{1}}]+tr_{2}[\rho _{A_{2}}\log
_{2}\sigma _{A_{2}}].
\end{eqnarray}
It is known that $\rho _{A_{1}A_{2}}$ can be written as \cite{Schlienz1995}
\begin{eqnarray}
\rho _{A_{1}A_{2}}=\sum_{j}c_{j}\rho _{1j}\otimes \rho _{2j},
\end{eqnarray}
where $\{c_{j}\}_{j}$ are real numbers, $\rho _{1j}$, $\rho _{2j}$ are all
Hermite matrices. For $\sigma _{A_{1}}$, $\sigma _{A_{2}}$, there exist
unitary matrices $U_{1}$ and $U_{2}$ such that $D_{1}=U_{1}\sigma
_{A_{1}}U_{1}^{+}$, $D_{2}=U_{2}\sigma _{A_{2}}U_{2}^{+}$ are all diagonal,
where $+$ denotes adjoint. Note that
\begin{eqnarray}
\log _{2}(D_{1}\otimes D_{2})=(\log _{2}D_{1})\otimes I_{2}+I_{1}\otimes (\log
_{2}D_{2}),
\end{eqnarray}
where $I_{1}$, $I_{2}$ are the identity operators on $H_{1}$, $H_{2}$,
respectively. Then Eq.(11) can be directly verified. $\Box$

With the help of Lemma 1 and Lemma 2, we can get an equivalent expression
for GQD defined by Eq.(7).

\emph{Theorem 1}. The GQD of a state $\rho _{A_{1}A_{2}...A_{N}}$ defined by Eq.(7)
can also be expressed as
\begin{eqnarray}
D(\rho _{A_{1}A_{2}...A_{N}})  \ \ \ \ \ \ \ \ \ \ \  \ \ \ \ \ \ \ \ \ \ \  \ \ \ \ \ \ \ \ \ \ \  \ \ \ \ \ \ \ \ \ \ \    \nonumber \\
=\min_{\Phi }[I(\rho
_{A_{1}A_{2}...A_{N}})-I(\Phi _{A_{1}A_{2}...A_{N}}(\rho
_{A_{1}A_{2}...A_{N}}))],
\end{eqnarray}
where, the mutual information
\begin{eqnarray}
I(\rho _{A_{1}A_{2}...A_{N}})=\sum_{i=1}^{N}S(\rho _{A_{i}})-S(\rho
_{A_{1}A_{2}...A_{N}}).
\end{eqnarray}
\emph{Proof}. From lemma 1, we have
\begin{eqnarray}
S(\rho _{A_{j}}||\Phi _{A_{j}}(\rho _{A_{j}}))=-S(\rho _{A_{j}})+S(\Phi
_{A_{j}}(\rho _{A_{j}})).
\end{eqnarray}
Note that
\begin{eqnarray}
I(\rho _{A_{1}A_{2}...A_{N}})=\sum_{i=1}^{N}S(\rho _{A_{i}})-S(\rho
_{A_{1}A_{2}...A_{N}})        \nonumber \\
=S(\rho _{A_{1}A_{2}...A_{N}}||\rho _{A_{1}}\otimes \rho _{A_{2}}\otimes
...\otimes \rho _{A_{N}}).
\end{eqnarray}
Together with Lemma 2, we can easily prove Theorem 1. $\Box$

From Theorem 1, we see that, GQD of a state is just the minimal loss of mutual
information over all locally projective measurements. This interpretation of
GQD is consistency with the symmetric quantum discord in Eq.(1), as well as
the original definition of quantum discord for bipartite states.

For a special case, we consider a state $\rho _{A_{1}A_{2}...A_{N}}$ whose
reduced states $\rho _{A_{1}}$, $\rho _{A_{2}}$, ... , $\rho _{A_{N}}$ are
all proportional to identity operator. In such case, the GQD of $\rho
_{A_{1}A_{2}...A_{N}}$ can be remarkably simplified. We state it as Theorem
2, its proof is easy.

\emph{Theorem 2}. An $N$-partite state $\rho _{A_{1}A_{2}...A_{N}}$, if its reduced
states $\rho _{A_{1}}$, $\rho _{A_{2}}$, ... , $\rho _{A_{N}}$ are all
proportional to identity operator, then the GQD of $\rho
_{A_{1}A_{2}...A_{N}}$ can be expressed as
\begin{eqnarray}
D(\rho _{A_{1}A_{2}...A_{N}})=-S(\rho _{A_{1}A_{2}...A_{N}})   \ \ \ \ \ \ \ \ \ \ \ \ \ \ \ \ \  \ \ \ \    \nonumber \\
+\min_{\Phi}S(\Phi _{A_{1}A_{2}...A_{N}}(\rho _{A_{1}A_{2}...A_{N}})).
\end{eqnarray}

\section{two classes of $N$-qubit states which allow analytical expressions of GQD}

We consider two classes of $N$-qubit states which allow analytical
expressions of GQD. We first recall two mathematical facts.

\emph{Lemma 3}. \cite{Cornwell1984} Group homomorphism of U(2) to SO(3).

For any two-dimensional unitary matrix $U$, there exists a unique real
three-dimensional orthogonal matrix $R$ with determinant 1, such that for
any real three-dimensional vector $\vec{r}=(r_{x},r_{y},r_{z})$ ,
it holds that
\begin{eqnarray}
U\vec{r}\cdot \vec{\sigma }U^{+}=R(\vec{r}%
)\cdot \vec{\sigma }.
\end{eqnarray}
Conversely, For any real three-dimensional orthogonal matrix $R$ with
determinant 1, there exists (not unique) a two-dimensional unitary matrix $U$%
, fulfills Eq.(19).

In Eq.(19), $\vec{r}\cdot \vec{\sigma }=r_{x}\sigma
_{x}+r_{y}\sigma _{y}+r_{z}\sigma _{z}$, $\vec{\sigma }=\{\sigma
_{x},\sigma _{y},\sigma _{z}\}$ are Pauli matrices; $R(\vec{r})=(R%
\vec{r}^{t})^{t}$ is a real three-dimensional vector, here $t$
denotes matrix transpose.

\emph{Lemma 4}. \cite{Marshal1979} Monotonicity of entropy function under majorization relation.

For given $\{p_{1},p_{2},...,p_{n}\}$, $\{q_{1},q_{2},...,q_{n}\}$, satisfy

$1\geq p_{1}\geq p_{2}\geq ...\geq p_{n}\geq 0$, $\ \sum_{i=1}^{n}p_{i}=1$,

$1\geq q_{1}\geq q_{2}\geq ...\geq q_{n}\geq 0$, $\ \sum_{i=1}^{n}q_{i}=1$.
If
\begin{eqnarray}
\sum_{i=1}^{k}p_{i}\leq \sum_{i=1}^{k}q_{i},k=1,2,...,n,
\end{eqnarray}
then
\begin{eqnarray}
-\sum_{i=1}^{n}p_{i}\log _{2}p_{i}\geq -\sum_{i=1}^{n}q_{i}\log _{2}q_{i}.
\end{eqnarray}

Theorem 3 and Theorem 4 below provide two classes of multi-qubit states
which allow analytical expressions of GQD.

\emph{Theorem 3}. For $N$-qubit $(N\geq 2)$ Werner-GHZ state
\begin{eqnarray}
\rho =(1-\mu )\frac{I^{\otimes N}}{2^{N}}+\mu |\psi \rangle \langle \psi |,
\end{eqnarray}
the GQD of $\rho $ is
\begin{gather}
D(\rho )=(\frac{1-\mu }{2^{N}}+\mu )\log _{2}(\frac{1-\mu }{2^{N}}+\mu )+
\frac{1-\mu }{2^{N}}\log _{2}\frac{1-\mu }{2^{N}}    \nonumber  \\
-2(\frac{1-\mu }{2^{N}}+\frac{\mu }{2})\log _{2}(\frac{1-\mu }{2^{N}}+\frac{%
\mu }{2}).  \ \ \ \ \ \
\end{gather}
Where, I is $2\times 2$ identity operator, $\mu \in \lbrack 0,1]$, $|\psi
\rangle $ is the $N$-qubit GHZ state
\begin{eqnarray}
|\psi \rangle =(|00...0\rangle +|11...1\rangle )/\sqrt{2}.
\end{eqnarray}

\emph{Proof}. A projective measurement on single qubit can be expressed as
\begin{eqnarray}
\Pi _{0}=\frac{1}{2}(I+\vec{\Pi }\cdot \vec{\sigma }%
),   \ \
\Pi _{1}=\frac{1}{2}(I-\vec{\Pi }\cdot \vec{\sigma }%
),
\end{eqnarray}
where, $\vec{\Pi }=(\alpha ,\beta ,\gamma )$ is a real vector
with unit length.

It is easy to check that
\begin{gather}
\vec{\Pi }(\sigma _{x})=\Pi _{0}\sigma _{x}\Pi _{0}+\Pi
_{1}\sigma _{x}\Pi _{1}=\alpha \vec{\Pi }\cdot \vec{%
\sigma },  \\
\vec{\Pi }(\sigma _{y})=\Pi _{0}\sigma _{y}\Pi _{0}+\Pi
_{1}\sigma _{y}\Pi _{1}=\beta \vec{\Pi }\cdot \vec{%
\sigma },   \\
\vec{\Pi }(\sigma _{z})=\Pi _{0}\sigma _{z}\Pi _{0}+\Pi
_{1}\sigma _{z}\Pi _{1}=\gamma \vec{\Pi }\cdot \vec{%
\sigma }.
\end{gather}
Now for any locally projective measurement performing on $N$-qubit state $%
\rho $, we label it by $\vec{\Pi }_{1}=(\alpha _{1},\beta
_{1},\gamma _{1})$, $\vec{\Pi }_{2}=(\alpha _{2},\beta
_{2},\gamma _{2})$, ... , $\vec{\Pi }_{N}=(\alpha _{N},\beta
_{N},\gamma _{N})$, each of them is a real vector with unit length. It is
easy to verify that for state $\rho $ in Eq.(22), whose reduced states $\rho
_{A_{1}}$, $\rho _{A_{2}}$, ... , $\rho _{A_{N}}$ are all proportional to
identity operator. So, we can calculate $D(\rho )$ according to Theorem 2.
We then need to calculate $S(\rho )$ and $S(\Phi _{A_{1}A_{2}...A_{N}}(\rho
_{A_{1}A_{2}...A_{N}})).$

The eigenvalues of $\rho $ can be easily found, that is
\begin{gather}
Spec(\rho )=\{\frac{1-\mu }{2^{N}}+\mu ,\frac{1-\mu }{2^{N}},\frac{1-\mu }{%
2^{N}},...,\frac{1-\mu }{2^{N}}\}.
\end{gather}
$S(\rho )$ can then be directly calculated by Eq.(29).

Let $\Phi _{A_{1}A_{2}...A_{N}}$ be the locally projective measurement $\{%
\vec{\Pi }_{i}\}_{i=1}^{N}$, using Eqs.(26-28) and the facts
\begin{gather}
|\psi \rangle \langle \psi |=\frac{1}{2}(|0\rangle \langle 0|^{\otimes
N}+|0\rangle \langle 1|^{\otimes N}   \nonumber  \\     \ \ \ \ \ \ \ \ \ \ \ \ \ \ \ \
+|1\rangle \langle 0|^{\otimes N}+|1\rangle \langle 1|^{\otimes N}),  \\
|0\rangle \langle 0|=\frac{I+\sigma _{3}}{2},  \ \ \
|1\rangle \langle 1|=\frac{I-\sigma _{3}}{2},   \\   \ \ \ \ \ \ \ \
|0\rangle \langle 1|=\frac{\sigma _{1}+i\sigma _{2}}{2},   \  \
|1\rangle \langle 0|=\frac{\sigma _{1}-i\sigma _{2}}{2},
\end{gather}
we get
\begin{gather}
\Phi _{A_{1}A_{2}...A_{N}}(|\psi \rangle \langle \psi |)    \ \ \ \ \ \ \ \ \ \ \ \ \ \ \ \ \ \ \ \ \ \ \ \ \ \ \ \ \ \ \ \ \ \ \ \  \nonumber \\
=\frac{1}{2}[\otimes _{i=1}^{N}\frac{I+\gamma _{i}\vec{\Pi }%
_{i}\cdot \vec{\sigma }}{2}+\otimes _{i=1}^{N}\frac{\alpha
_{i}+i\beta _{i}}{2}\vec{\Pi }_{i}\cdot \vec{\sigma }  \ \ \ \ \ \ \ \ \ \ \ \nonumber \\
+\otimes _{i=1}^{N}\frac{\alpha _{i}-i\beta _{i}}{2}\vec{\Pi }%
_{i}\cdot \vec{\sigma }+\otimes _{i=1}^{N}\frac{I-\gamma _{i}%
\vec{\Pi }_{i}\cdot \vec{\sigma }}{2}].
\end{gather}
For given $\{\overrightarrow{\Pi }_{i}\}_{i=1}^{N}$, according to Lemma 3,
we can always perform a locally unitary transformation on $\Phi
_{A_{1}A_{2}...A_{N}}(\rho )$ ($S(\Phi _{A_{1}A_{2}...A_{N}}(\rho ))$ keeps
invariant) such that $\vec{\Pi }_{i}\cdot \vec{\sigma }$ is transformed to
$\sigma _{z}$ for all $i$, then Eq.(33) becomes
\begin{gather}
\frac{1}{2}[\otimes _{i=1}^{N}\frac{I+\gamma _{i}\sigma _{z}}{2}+\otimes
_{i=1}^{N}\frac{I-\gamma _{i}\sigma _{z}}{2}  \ \ \ \ \ \ \      \nonumber \\
+\otimes _{i=1}^{N}\frac{\alpha _{i}+i\beta _{i}}{2}\sigma _{z}+\otimes
_{i=1}^{N}\frac{\alpha _{i}-i\beta _{i}}{2}\sigma _{z}].
\end{gather}
Eq.(34) is a diagonal matrix with $2^{N}$ diagonal elements
\begin{gather}
\frac{1}{2}[\prod _{i=1}^{N}\frac{1+(-1)^{n_{i}}\gamma _{i}}{2}+\prod _{i=1}^{N}%
\frac{I-(-1)^{n_{i}}\gamma _{i}}{2}     \ \ \ \ \ \ \ \    \nonumber \\
+\prod _{i=1}^{N}\frac{\alpha _{i}+(-1)^{n_{i}}i\beta _{i}}{2}+\prod _{i=1}^{N}%
\frac{\alpha _{i}-(-1)^{n_{i}}i\beta _{i}}{2}],
\end{gather}
where, $n_{1},n_{2},...,n_{N}\in \{0,1\}.$
Consequently, $\Phi _{A_{1}A_{2}...A_{N}}(\rho )$ have $2^{N}$ eigenvalues
\begin{gather}
\frac{1-\mu }{2^{N}}+\frac{\mu }{2}[\prod _{i=1}^{N}\frac{1+(-1)^{n_{i}}%
\gamma _{i}}{2}+\prod _{i=1}^{N}\frac{I-(-1)^{n_{i}}\gamma _{i}}{2}    \nonumber \\
+\prod _{i=1}^{N}\frac{\alpha _{i}+(-1)^{n_{i}}i\beta _{i}}{2}+\prod _{i=1}^{N}%
\frac{\alpha _{i}-(-1)^{n_{i}}i\beta _{i}}{2}],
\end{gather}
where, $n_{1},n_{2},...,n_{N}\in \{0,1\}.$

Note that in Eq.(36), the eigenvalues corresponding to $\{n_{1},n_{2},...,n_{N}\}$ and $\{1-n_{1},1-n_{2},...,1-n_{N}\}$ are equal. Together with Lemma 4, we then assert that when $\gamma _{i}=1$ for all $i$, $S(\Phi
_{A_{1}A_{2}...A_{N}}(\rho ))$ achieves its minimum, since when $\gamma _{i}=1$ for all $i$, $\Phi _{A_{1}A_{2}...A_{N}}(\rho )$ have $2^{N}$ eigenvalues
 \begin{gather}
 \{\frac{1-\mu }{2^{N}}+\frac{\mu }{2},\frac{1-\mu }{2^{N}}+\frac{\mu }{2},\frac{1-\mu }{2^{N}},\frac{1-\mu }{2^{N}},...,\frac{1-\mu }{2^{N}}\}.
 \end{gather}
 So, by Eq.(37) and Eq.(29), Theorem 3 can be proved. $\Box$

We consider the behavior of Eq.(23) with large number $N$. Let $N\rightarrow
\infty $, we get Corollary 1 below.

\emph{Corollary 1}. When $N\rightarrow \infty $, Eq.(23) approximates
 \begin{gather}
 D_{\infty}(\rho )=\mu.
 \end{gather}

We remark that, Eq.(38) is somehow a good approximation of Eq.(23). In fact, when
$N=10$, the deviation is less than $10^{-2}$; $N=14$, the deviation is less
than $10^{-3}$; $N=17$, the deviation is less than $10^{-4}$. Fig.1 shows
Eq.(23) for $N=2$, $3$, $5$, $\infty$, respectively.

We also remark that, when $N=3$, Eq.(23) is the same to the result obtained in \cite{Rulli2011}. When $N=2$%
, Eq.(23) returns the result of Werner state considered in \cite{Zurek2001}.

\begin{figure}[h]
\includegraphics[width=8.4cm]{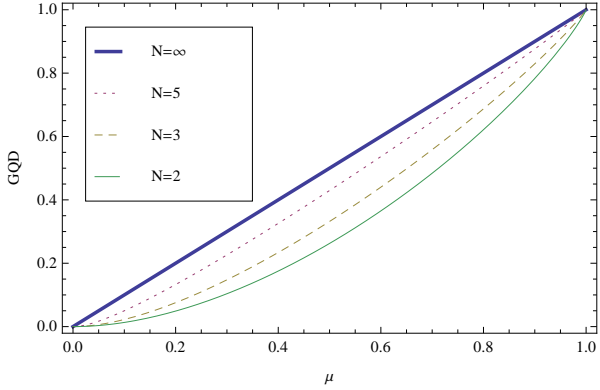}
\caption{GQD of Werner-GHZ state in Eq.(22) when $N=2$, $3$, $5$, $\infty$, respectively.}
\label{fig1}
\end{figure}

We next consider another class of $N$-qubit states.

\emph{Theorem 4}. For $N$-qubit state
 \begin{gather}
\rho =\frac{1}{2^{N}}(I^{\otimes N}+c_{1}\sigma _{x}^{\otimes
N}+c_{2}\sigma _{y}^{\otimes N}+c_{3}\sigma _{z}^{\otimes N}),
 \end{gather}
the GQD of $\rho $ is
 \begin{gather}
D(\rho )=f(\rho )-g(\rho ).
 \end{gather}
In Eq.(40),
\begin{gather}
f(\rho )=-\frac{1+c}{2}\log _{2}\frac{1+c}{2}-\frac{1-c}{2}\log _{2}\frac{%
1-c}{2},   \\
c=max\{|c_{1}|,|c_{2}|,|c_{3}|\};
\end{gather}
when $N$ is odd,
\begin{gather}
g(\rho )=-\frac{1+d}{2}\log _{2}\frac{1+d}{2}-\frac{1-d}{2}\log _{2}\frac{%
1-d}{2},  \\
d=\sqrt{c_{1}^{2}+c_{2}^{2}+c_{3}^{2}};
\end{gather}
when N is even,
\begin{gather}
g(\rho )=-1-\sum_{j=1}^{4}\lambda _{j}\log _{2}\lambda _{j},\\
\lambda _{1}=[1+c_{3}+c_{1}+(-1)^{N/2}c_{2}]/4,\\
\lambda _{2}=[1+c_{3}-c_{1}-(-1)^{N/2}c_{2}]/4,\\
\lambda _{3}=[1-c_{3}+c_{1}-(-1)^{N/2}c_{2}]/4,\\
\lambda _{4}=[1-c_{3}-c_{1}+(-1)^{N/2}c_{2}]/4.
\end{gather}
In Eq.(39), $I$ is the $2\times 2$ identity operator, $\{c_{1},c_{2},c_{3}\}$
are real numbers constrained by $d\in \lbrack 0,1]$ (when $N$ is odd) or $%
\lambda _{1}$, $\lambda _{2}$, $\lambda _{3}$, $\lambda _{4}\in \lbrack 0,1]$
(when $N$ is even).

\emph{Proof}. Obviously, for state $\rho $ in Eq.(39), its reduced states $\rho _{A_{1}}$%
, $\rho _{A_{2}}$, ... , $\rho _{A_{N}}$ are all proportional to identity
operator. Thus, we can calculate $D(\rho )$ according to Theorem 2. We then need to
calculate $S(\rho )$ and $S(\Phi _{A_{1}A_{2}...A_{N}}(\rho
_{A_{1}A_{2}...A_{N}})).$ It is easy to see that state $\rho $ has nonzero
elements only on the principle diagonal and the antidiagonal, so is the
matrix $(\rho -xI^{\otimes N})$. By the Laplace theorem in linear algebra, $%
M=det(\rho -xI^{\otimes N})$ can be expanded as the multiplications of $%
2^{N-1}$ determinants of $2\times 2$ matrices,
\begin{gather}
M=det(\rho -xI^{\otimes N})   \ \ \ \ \ \ \ \ \ \ \ \ \ \ \ \ \ \ \ \ \ \ \ \ \ \ \ \ \ \ \ \ \ \ \    \\
=\prod _{j=1}^{2^{N-1}}\det \left(
\begin{array}{cc}
M_{jj} & M_{j,2^{N}-j} \\
M_{2^{N}-j,j} & M_{2^{N}-j,2^{N}-j}%
\end{array}%
\right),
\end{gather}
where
\begin{gather}
\ \ \ \ \ \ \   M_{jj}=\frac{1}{2^{N}}+(-1)^{n_{1}+n_{2}+...+n_{N}}\frac{c_{3}}{2^{N}}-x,  \\
M_{2^{N}-j,2^{N}-j}=\frac{1}{2^{N}}+(-1)^{N+n_{1}+n_{2}+...+n_{N}}%
\frac{c_{3}}{2^{N}}-x,   \\
M_{2^{N}-j,j}=\frac{c_{1}}{2^{N}}+i^{N}(-1)^{n_{1}+n_{2}+...+n_{N}}\frac{%
c_{2}}{2^{N}},  \      \\
\ \ \   M_{j,2^{N}-j}=\frac{c_{1}}{2^{N}}+i^{N}(-1)^{N+n_{1}+n_{2}+...+n_{N}}%
\frac{c_{2}}{2^{N}},
\end{gather}
$\{n_{1},n_{2},...,n_{N}\}\in \{0,1\}$.

Then by direct calculation of $det(\rho -xI^{\otimes N})=0$, we get the
eigenvalues of $\rho $. That is, when $N$ is even, the eigenvalues of $\rho $
are
\begin{gather}
\{\frac{1+c_{3}}{2^{N}}\pm \frac{c_{1}+(-1)^{N/2}c_{2}}{2^{N}},\frac{1-c_{3}%
}{2^{N}}\pm \frac{c_{1}-(-1)^{N/2}c_{2}}{2^{N}}\},
\end{gather}
each of them possesses multiplicity $2^{N-2}$. When $N$ is odd, the
eigenvalues of $\rho $ are
\begin{gather}
\{\frac{1}{2^{N}}(1\pm \sqrt{c_{1}^{2}+c_{2}^{2}+c_{3}^{2}})\},
\end{gather}
each of them possesses multiplicity $2^{N-1}$.

$S(\rho )$ can then be directly calculated by Eqs.(56-57).

We now calculate $\Phi _{A_{1}A_{2}...A_{N}}(\rho )$. After the measurement $\{\vec{\Pi }_{i}=(\alpha _{i},\beta
_{i},\gamma _{i})\}_{i=1}^{N}$ as labelled in the proof of Theorem 3, from Eqs.(26-28), we have
\begin{gather}
\Phi _{A_{1}A_{2}...A_{N}}(\rho )=\frac{1}{2^{N}}[I^{\otimes N}
+(c_{1}\prod _{i=1}^{N}\alpha _{i}  \ \ \ \ \ \ \ \ \ \ \ \ \ \ \ \ \ \ \ \ \     \nonumber  \\   \ \ \ \ \ \ \ \ \ \ \ \
+c_{2}\prod_{i=1}^{N}\beta _{i}+c_{3}\prod _{i=1}^{N}\gamma _{i})\otimes _{i=1}^{N}(%
\vec{\Pi }_{i}\cdot \vec{\sigma })].
\end{gather}
From Lemma 3, for any $\{\vec{\Pi }_{i}\}_{i=1}^{N}$ we can
always perform a locally unitary transformation on $\Phi
_{A_{1}A_{2}...A_{N}}(\rho )$ ($S(\Phi _{A_{1}A_{2}...A_{N}}(\rho ))$ keeps
invariant) such that $\vec{\Pi }_{i}\cdot \vec{\sigma }$ is transformed to
$\sigma _{z}$ for all $i$, then Eq.(58) becomes
\begin{gather}
\frac{1}{2^{N}}[I^{\otimes N}+(c_{1}\prod _{i=1}^{N}\alpha _{i}+c_{2}\prod
_{i=1}^{N}\beta _{i}+c_{3}\prod _{i=1}^{N}\gamma _{i})\sigma _{z}^{\otimes
N}].
\end{gather}
Eq.(59) is a diagonal matrix, so its eigenvalues are
\begin{gather}
\{\frac{1}{2^{N}}[1\pm (c_{1}\prod _{i=1}^{N}\alpha _{i}+c_{2}\prod
_{i=1}^{N}\beta _{i}+c_{3}\prod _{i=1}^{N}\gamma _{i})]\},
\end{gather}
each of them has multiplicity $2^{N-1}$.

From Lemma 4, we find that minimizing $S(\Phi _{A_{1}A_{2}...A_{N}}(\rho ))$
is equivalent to maximizing
\begin{gather}
|c_{1}\prod _{i=1}^{N}\alpha _{i}+c_{2}\prod _{i=1}^{N}\beta _{i}+c_{3}\prod
_{i=1}^{N}\gamma _{i}|,
\end{gather}
over all possible $\{\vec{\Pi }_{i}\}_{i=1}^{N}$.

Suppose $\{\vec{\Pi }_{i}\}_{i=1}^{N-1}$ are given, then Eq.(61)
can be written as
\begin{gather}
|(\alpha _{N},\beta _{N},\gamma _{N})\cdot (c_{1}\prod _{i=1}^{N-1}\alpha
_{i},c_{2}\prod _{i=1}^{N-1}\beta _{i},c_{3}\prod _{i=1}^{N-1}\gamma _{i})|.
\end{gather}
So over all possible $\vec{\Pi }_{N}=(\alpha _{N},\beta _{N},\gamma _{N})$ with
$\alpha _{N}^{2}+\beta _{N}^{2}+\gamma _{N}^{2}=1$, Eq.(62) achieves the maximum
\begin{gather}
(c_{1}^{2}\prod _{i=1}^{N-1}\alpha _{i}^{2}+c_{2}^{2}\prod _{i=1}^{N-1}\beta
_{i}^{2}+c_{3}^{2}\prod _{i=1}^{N-1}\gamma _{i}^{2})^{1/2}.
\end{gather}
Suppose $\{\vec{\Pi }_{i}\}_{i=1}^{N-2}$ are given, because $\alpha _{N-1}^{2}+\beta _{N-1}^{2}+\gamma _{N-1}^{2}=1$, so the maximum of Eq.(63) over all possible $\vec{\Pi }_{N-1}=(\alpha _{N-1},\beta _{N-1},\gamma _{N})$ is
\begin{gather}
(max\{c_{1}^{2}\prod _{i=1}^{N-2}\alpha _{i}^{2},c_{2}^{2}\prod
_{i=1}^{N-2}\beta _{i}^{2},c_{3}^{2}\prod _{i=1}^{N-2}\gamma _{i}^{2}\})^{1/2}.
\end{gather}
The maximum of Eq.(64) over all possible $\{\vec{\Pi }_{i}\}_{i=1}^{N-2}$ apparently is
\begin{gather}
c=max\{|c_{1}|,|c_{2}|,|c_{3}|\}.
\end{gather}
By Eqs.(56,57,60,65), we then complete this proof. $\Box$

We remark that, when $N=2$, Theorem 4 recovers the result of 2-qubit
Bell-diagonal state which was first obtained in \cite{Luo2009}.

We know the original quantum discord may manifest the phenomena of sudden
transition and freeze \cite{Maziero2009,Mazzola2010,Guo2010}. With the analytical result of Theorem 4, we assert that  GQD can also manifest such interesting phenomena. We make this assertion more clear by giving an example. Let $N$ qubits be in the state Eq.(39), and let any qubit undergo a local phase damping
\begin{gather}
E_{0}=\sqrt{1-p/2} \ I,  \ \ \ \ \  E_{1}=\sqrt{p/2} \ \sigma _{z}.
\end{gather}
 After this channel, the state $\rho $ in Eq.(39) becomes
\begin{gather}
\rho (p)=\frac{1}{2^{N}}(I^{\otimes N}+c_{1}(p)\sigma _{x}^{\otimes
N}+c_{2}(p)\sigma _{y}^{\otimes N}+c_{3}(p)\sigma _{z}^{\otimes N}),
\end{gather}
where
\begin{gather}
c_{1}(p)=c_{1}(1-p), \
c_{2}(p)=c_{2}(1-p), \
c_{3}(p)=c_{3}.
\end{gather}
So, $D(\rho (p))$ can be calculated by Theorem 4. 
Therefore, similar to the bipartite case discussed in \cite{Maziero2009,Mazzola2010}, it can be found
that, the sudden transition occurs if and only if
\begin{gather}
0<|c_{3}|<\max \{|c_{1}|,|c_{2}|\},
\end{gather}
and freezing GQD may occur when N is even.

\section{conclusion}

 In summary, we provided an equivalent expression of global quantum
discord (GQD). From this equivalent expression, we gave an interpretation of
GQD as the minimal loss of mutual information over all locally projective
measurements. This interpretation is consistency with the original quantum
discord. We obtained the analytical expressions of GQD for two classes of
multi-qubit states, each of them possesses high symmetry. \ By the
analytical expressions of these states, we discussed some behaviors of GQD,
including the asymptotic behavior of GQD when $N$ tends infinity, and the
phenomena of sudden transition and freeze of GQD.

This work was supported by the Fundamental Research Funds for the Central Universities of
China (Grant No.2010scu23002). The author thanks Qing Hou and Bo You for helpful discussions.

\end{document}